\documentclass[12pt,preprint]{emulateapj}
\usepackage{graphicx}

\shorttitle{A Necro-Biological Explanation for the Fermi Paradox}
\shortauthors{Stephen R. Kane \& Franck Selsis}
\slugcomment{Submitted for publication in the Necronomicon}

\begin{document}

\title{A Necro-Biological Explanation for the Fermi Paradox}
\author{
  Stephen R. Kane\altaffilmark{1} \&
  Franck Selsis\altaffilmark{2}
}
\affiliation{$^1$ Center for Global Extinction Pandemic Control,
  Subterranean Bunker 32, Union Square, San Francisco, USA \\
$^2$ Planetary Defense Institute - Zombie Division, Chateau
  Morts-Vivants, Bordeaux, France}


\begin{abstract}

As we learn more about the frequency and size distribution of
exoplanets, we are discovering that terrestrial planets are
exceedingly common. The distribution of orbital periods in turn
results in many of these planets being the occupants of the Habitable
Zone of their host stars. Here we show that a conclusion of prevalent
life in the universe presents a serious danger due to the risk of
spreading Spontaneous Necro-Animation Psychosis (SNAP), or
Zombie-ism. We quantify the extent of the danger posed to Earth
through the use of the Zombie Drake Equation and show how this serves
as a possible explanation for the Fermi Paradox. We demonstrate how to
identify the resulting necro-signatures present in the atmospheres
where a zombie apocalypse may have occurred so that the risk may be
quantified. We further argue that it is a matter of planetary defense
and security that we carefully monitor and catalog potential
SNAP-contaminated planets in order to exclude contact with these
worlds in a future space-faring era.

\end{abstract}



\section{Introduction}
\label{intro}

The detection of planets outside of our Solar System has opened up the
possibility of answering several questions which have nagged the minds
of philosophers for millennia. These questions include: Is the
architecture of our Solar System typical or unusual? How common are
planets the size of the Earth? How common is life in the universe?
Exactly how many things are out there that can kill us? It is now
apparent that the process of planet formation produces an enormous
diversity of planetary systems. It is also clear from more recent
discoveries, most notably those from the {\em Kepler} mission, that
terrestrial-size planets are exceptionally common. The primary
motivation for establishing such a correlation lies within the search
for life exterior to our Solar System and thus determine if life is
common. The fact that Earth-size planets are relatively common is
surely good news for resolving this issue. This may be true, but a
positive deduction that life is common may have a serious negative
consequence.

The evolution of life on Earth has been accompanied by symbiotic
relationships between animal species and the bacteria and viruses
which use the animals as hosts. These occasionally result in
destructive outcomes which have had a devastating impact on various
populations of animals due to genetic breakdowns caused by the
virus. Particularly lethal pandemics which have affected homo-sapiens
in recent centuries include cholera, influenza, typhus, and smallpox.
A more recent phenomenon which has been studied in great detail is
that of Spontaneous Necro-Animation Psychosis (SNAP), often referred
to as Zombie-ism. This highly contagious condition is particularly
nefarious in so far as its use of the host itself to provide a mobile
platform from which to consciously spread the condition. Detailed
modeling of various SNAP outbreak scenarios by \citet{mun09} have
shown that human civilization would not only be unlikely to survive
such an event but would collapse remarkably quickly.

Here we discuss how recent exoplanet discoveries combined with studies
of infectious diseases indicate that the universe may harbor
reservoirs of planets full of bio-decay remains where zombie
apocalypses have occurred. In Section \ref{danger} we outline the
dangerous nature of SNAP, quantify the possible numbers of
SNAP-contaminated planets, and their proximity to Earth. In Section
\ref{decomp} we describe the decomposition process and the gases
released. This process is then used to establish the resulting
necro-signatures and their potential for identification in Section
\ref{necro}. The observing window for detecting such signatures is
discussed in Section \ref{window} and we provide the final sobering
and terrifying conclusions in Section \ref{conclusions}.


\section{The Reality of the Danger}
\label{danger}

Spontaneous Necro-Animation Psychosis is undoubtedly the most
dangerous viral condition to infect living organisms. The infectious
nature of the condition is maximized by bestowing upon the host an
insatiable desire to spread the virus at all costs. This ensures that
it will spread quickly and, usually, uncontrollably.

Although there have not yet been documented cases of SNAP outbreaks on
Earth, the reality of the condition has been extensively depicted in
both literature and cinema \citep{rus05,vuc11,kay12}. The science of
zombie-ism has been investigated and found a SNAP outbreak could
equally result from both natural evolution and genetic engineering
\citep{swa13}. In either case, defense from such an outbreak has also
been explored in great detail to maximize the survival probability
\citep{bro03,bro09}. Even with such defenses, the global scale of the
outbreak will rapidly break down any existing civilization. The novel
``World War Z'' depicts one such scenario although it presents an
unlikely end result in which humans are able to recover, albeit at the
brink of extinction \citep{bro07}. The work of \citet{mun09} more
accurately quantifies the likely outcome of a complete extinction
event occurring.

Although detecting the necro-signatures of worlds where a zombie
apocalypse has occurred, we can also estimate the number of worlds
which are affected in this way. To accomplish this, we use a modified
version of the well-known {\em Drake Equation}, the original of which
takes the following form:
\begin{equation}
  N = R_\star \times f_p \times n_e \times f_l \times f_i \times f_c
  \times L
  \label{drake}
\end{equation}
where the purpose is to calculate $N$ which is the number of advanced
civilizations in the galaxy with radio-communication capability. The
other variables include $R_\star$ (average rate of star formation),
$f_p$ (fraction of stars with planets), $n_e$ (average number of
life-capable planets per star), $f_l$ (fraction of planets with life),
$f_i$ (fraction of planets with intelligent life), $f_c$ (fraction of
civilizations that develop radio technology), and $L$ (length of time
such civilizations communicate). For a zombie outbreak to occur, there
is no reason to {\em a priori} assume that intelligent life is
required. Thus the modified {\em Zombie Drake Equation} is as follows:
\begin{equation}
  N_z = R_\star \times f_p \times n_e \times f_l \times f_z
  \label{moddrake}
\end{equation}
where $N_z$ is the total number of SNAP-contaminated planets and $f_z$
is the fraction of planets where an outbreak has utterly destroyed the
local population.

\begin{figure}
  \includegraphics[angle=270,width=8.2cm]{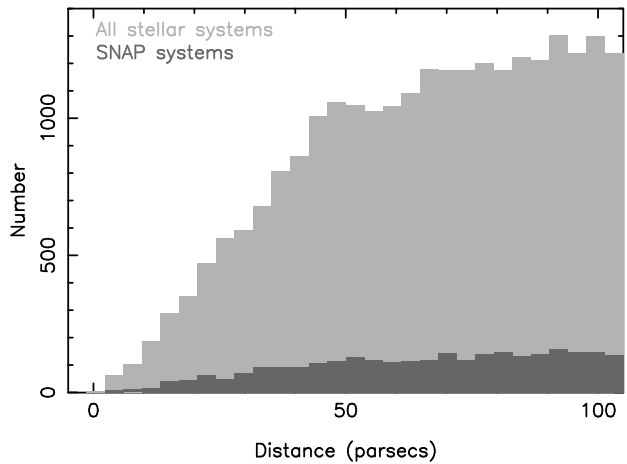}
  \caption{Histogram of all stars within 100 parsecs of the Earth
    (light gray) and the systems which are likely to contain
    SNAP-infected planets (dark gray). This shows that there are
    likely more than 2,500 SNAP-contaminated systems in close
    proximity to the Solar System.}
  \label{planets}
\end{figure}

\interfootnotelinepenalty=10000

There is strong evidence to suggest that the appearance of Earth-based
life occurred at a very early stage of Earth's history, probably at
least as early as 3.9 billion years ago. This lends credence to the
hypothesis that life is indeed a natural consequence of having
suitable conditions, which is a terrestrial planet within the
Habitable Zone of the star. Based on reasonable estimates of the
frequency of terrestrial planets with such conditions, we use Equation
\ref{moddrake} with a conservative $f_z$ value of 10\%. Shown in
Figure \ref{planets} is a histogram of all stars within 100 parsecs of
Earth based on data from the {\em Hipparcos} mission. The dark region
shows an estimate of the distribution of nearby stars which could have
a SNAP-contaminated planet based on the above assumptions. This would
mean that there are more than 2,500 such systems within 100 parsecs of
Earth. If that doesn't scare the bejeezus out of you then you may need
to check your pulse!\footnote{If there's no pulse, you could be our
  SNAP patient zero. The fact that you're reading this makes it
  unlikely. However, just to be sure, test if you can speak an
  intelligible sentence.}

Furthermore, the projected frequency of SNAP planets explains a
contradiction which has long troubled the proposition that intelligent
life is common: the {\em Fermi Paradox}. This premise of the paradox
is that the timescale for extraterrestrial civilizations to spread
throughout the galaxy is small compared with stellar lifetimes and so
we should have encountered our neighbors by now. Our work here shows
the resolution of the paradox to be quite simple. The desolation of a
civilization requires only that they encounter a case of SNAP during
their exploration phase and their entire civilization will
collapse. Let us not repeat history by rushing in to where our
predecessors ought to have feared to tread.


\section{Decomposition}
\label{decomp}

Now that your trousers are presumably the appropriate shade of brown,
we must determine how the SNAP planets may be detected remotely and
thus avoid them like ... well, the plague. To quantify this, we
first need to identify what it is we are actually looking for. A
defining outcome of a zombie apocalypse is the death of all animal
life on the planetary surface. This will result in the transfer of a
substantial fraction of the total biomass to the atmosphere through
the process of decomposition.

All animals undergo similar stages of decomposition: Fresh, Bloat,
Active Decay, Advanced Decay, and Dry remains. The primary
decomposition stages during which the purging of gases and fluids
occur are the Bloat and Active Decay stages. For Earth-based animals,
the primary gases produced during this process are carbon dioxide,
hydrogen sulfide, ammonia, and methane. The extent to which this
translates into strong signatures within the planetary atmosphere
depends on the relative mass of the biosphere being converted to these
gases. Humans constitute roughly 350 million tonnes of terrestrial
biomass with an additional 700 million tonnes available via
domesticated animals. The cross contamination of SNAP for humans
versus animals varies depending on the movie/literature source. Even
if humans are the only species to succumb to such an infection, we can
expect the levels of decomposition gases described above to at least
double and probably increase by factors of several.

Incidentally, the animal species on Earth with the highest biomass is
antarctic krill who apparently have natural selection all figured
out. However, it's unknown what a zombie apocalypse involving krill
would look like and there is certainly a cinema niche awaiting any
film director who would care to portray it.


\section{Detecting Necro-Signatures}
\label{necro}

\begin{figure*}
  \begin{center}
    \begin{tabular}{cc}
      \includegraphics[width=8.2cm]{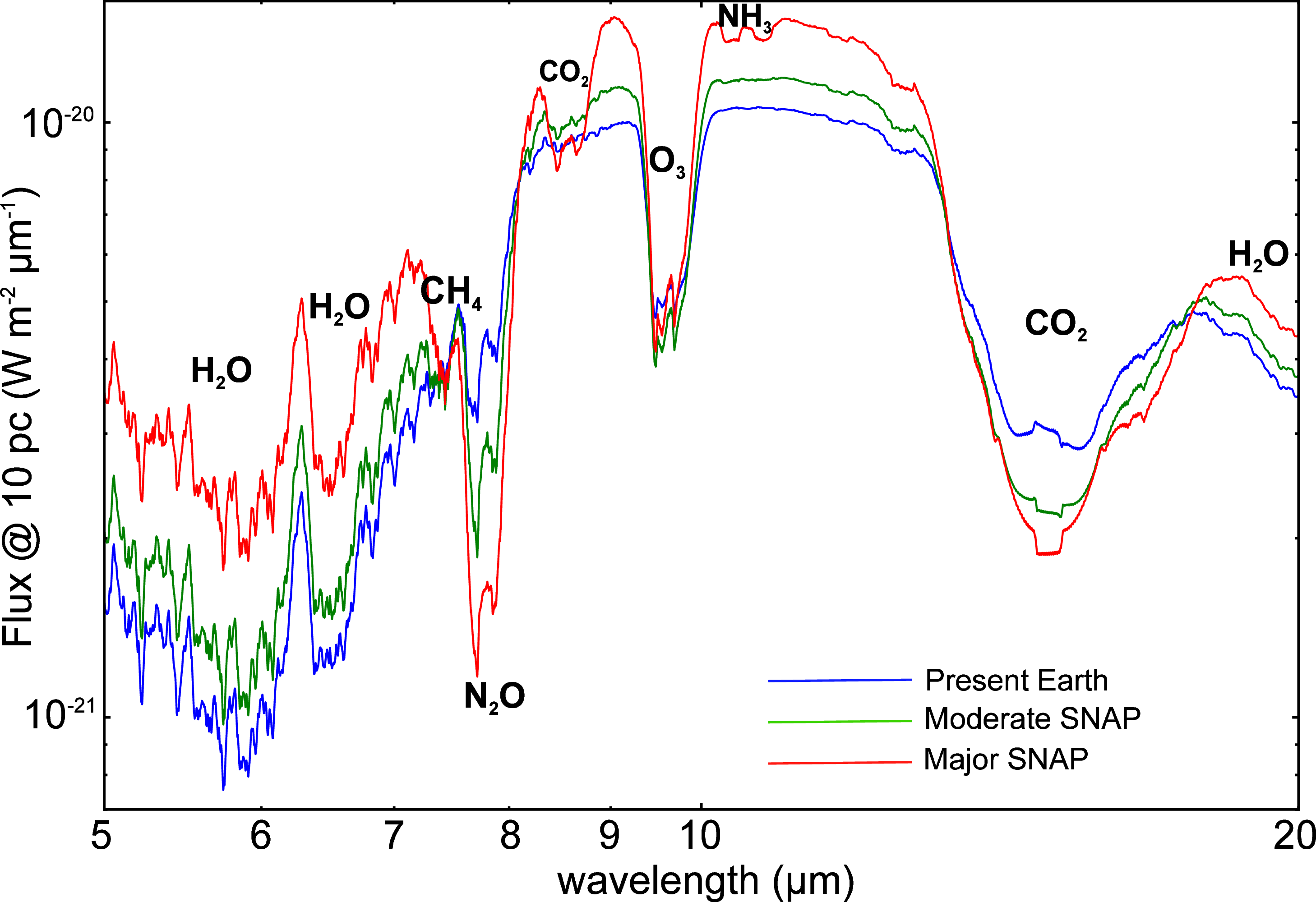} &
      \includegraphics[width=8.2cm]{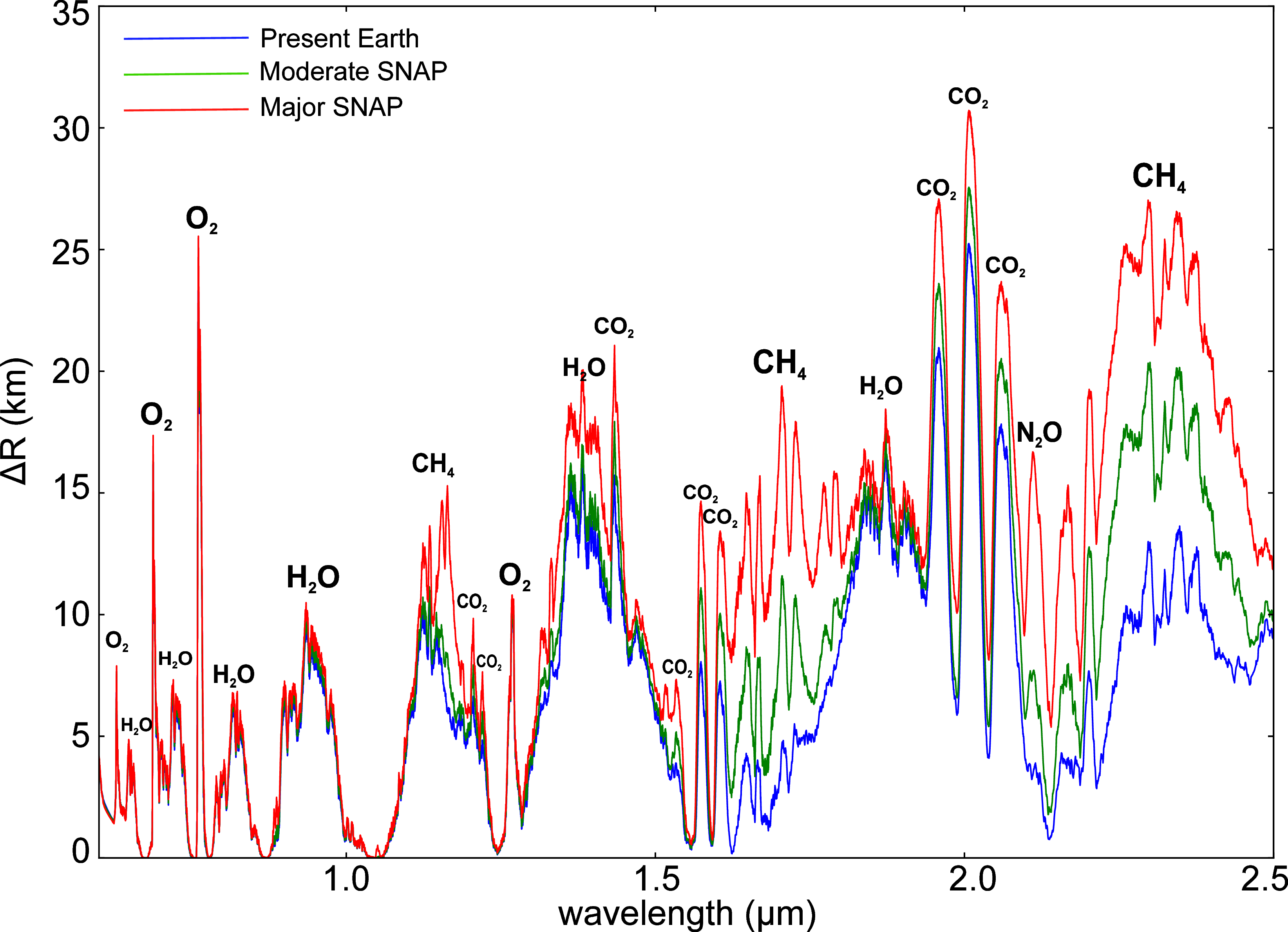}
    \end{tabular}
  \end{center}
  \caption{Thermal emission spectrum and transmission spectrum of an
    Earth-analog affected by SNAP. Left: Thermal emission as received
    at 10 pc. Right: Variation in the apparent radius given by the
    transit depth as a function of wavelength. The blue curves are for
    the present Earth. The green and red curves are for a moderate and
    major infection of the biosphere. See text for the corresponding
    atmospheric properties. Spectra computed assuming cloud-free
    conditions, with the code PAS (Post-Apocalyptic Spectroscopy,
    courtesy of Marcelino Agundez, Madrid).}
  \label{spectrum}
\end{figure*}

The gases released by the decomposition process described above may be
used to remotely detect the zombie afflicted worlds. In particular,
the strong atmospheric presence of CO$_2$, H$_2$S, CH$_4$, and NH$_3$
will reveal those locations where massive amounts of death and decay
have recently taken place. The strength of the respective signatures
for these gases in emission and transmission spectra will vary
greatly. The presence of increased levels of H$_2$S for example will
have a relatively weak associated signature. However, there are other
atmospheric processes that occur as part of apocalypse-level
decomposition that more than compensates and delivers an unambiguous
necro-signature.

Figure \ref{spectrum} shows the spectral signatures associated with a
zombie apocalypse on an Earth-like planet. The left panel shows the
transmission spectrum that can be obtained by transit spectroscopy
while the right panel shows the thermal emission of the planet that
can be observed either by secondary eclipse spectroscopy or infrared
nulling interferometry if the planet does not transit its host
star. Two cases are compared to present Earth. One is a moderate SNAP
with less than 10\% of an Earth-sized human and animal biomass
affected by the infection. This results in enhanced levels of some
atmospheric gases: 2 PAL (Present Atmospheric Level) of CO$_2$, and 5
PAL of CH$_4$, N$_2$O and NH$_3$, due to the putrefaction and
disruption of the nitrogen cycle (known as the Savini effect). This
also results in an increased greenhouse warming and a mean surface
temperature of 296~K instead of 288~K on present Earth. We also
modeled a major SNAP event, assuming a biomass twice the one on
present Earth and a 90\% infection. In this case, we find 4 PAL of
CO$_2$, 20 PAL of CH$_4$,and 50 PAL N$_2$O and NH$_3$. Spectral
features associated with all these species reveal the rise of zombies
at a planetary scale.

If you detect a signature such as the one described above, here is
what you should do. First, hyper-ventilate into a paper bag. Second,
call the person who occupies your country's highest office whilst
screaming hysterically. Neither action will help the situation but it
will make you feel like you've actually done something useful.


\section{Observing Window}
\label{window}

Due to the animation aspect of a zombie and its desire to infect
others, the corpse is invariably exposed to the
elements. Additionally, conflict between the zombies and those not yet
infected will produce high temperature conditions. The combination of
these two environmental effects will be to accelerate the rate of
decomposition (see Section \ref{decomp}) and thus produce a relatively
brief window in which signatures of the apocalypse may persist in the
atmosphere.

\begin{figure}
  \includegraphics[angle=270,width=8.2cm]{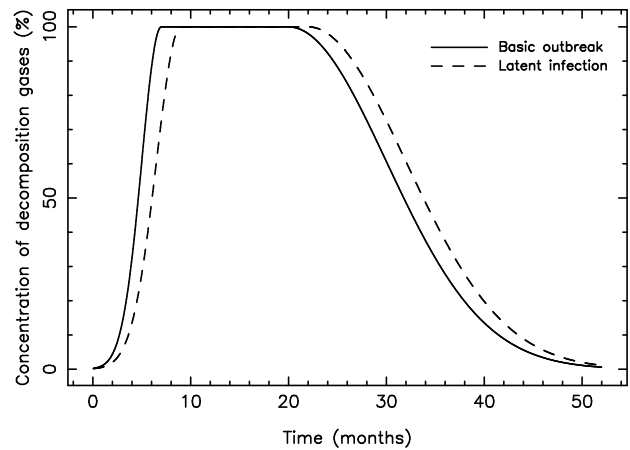}
  \caption{The concentration of decomposition gases as a function of
    time for both ``basic outbreak'' and ``latent infection''
    scenarios. For an Earth-size biomass, necro-signatures will remain
    at their peak amplitude for at least a year.}
  \label{conc}
\end{figure}

There are various models which may be used to determine the spread of
infection and thus the rise of decomposition gases in the atmosphere.
A broadly applicable model to use for the infection rate is the basic
model outbreak scenario of \citet{mun09} since we are not assuming
that the primary species with the infection is intelligent (yes,
including Earth). This model predicts an unimpeded spread of the
infection and a correspondingly rapid rise in decomposition
gases. This is shown by the solid line in Figure
\ref{conc}. Alternatively, one may consider the latent infection model
in which a certain fraction of the zombies are destroyed as they are
created by the astute uninfected population. As shown by the dashed
line in Figure \ref{conc}, this slows the release of decomposition
gases into the atmosphere but results in only a small delay. In either
case, the 100\% fatality rate produces a period of at least one year
during which the necro-signatures described in Section \ref{necro}
will be at their maximum amplitude, gradually decreasing over the
following couple of years. For biomasses larger than that currently
present on Earth, this period of maximum amplitude will be
proportionally longer. Removal of these gases from the atmosphere
assumes absorption by liquid water oceans and other mineral chemical
reactions. However, it is possible that a new equilibrium is reached
by which the necro-signatures could persist for much longer.


\section{Conclusions}
\label{conclusions}

We have shown that there is a significantly non-zero probability that
in the search for life in the universe we will also encounter large
amounts of undeath. Any person who has been exposed to even a
relatively benign zombie film understands the threat posed by this
heinous malady. This is not to be trifled with. Therefore the risk
imposed of encountering a SNAP-contaminated planet cannot be
overstated.

We have shown that the sign-posts for SNAP worlds are present and
detectable in exoplanet atmospheres. We have also shown that these
signatures may not persist for very long in the upper atmosphere which
emphasizes the need for continuous observations. An extension of the
necro-signature would be produced by worlds where advanced
civilizations existed due to the considerable time required for the
breakdown of the industry infrastructure left behind. One may well
point out that there are numerous scenarios other than a zombie
apocalypse that could equally quell all life on a planet. However, we
argue that none of those scenarios are anywhere near as scary as being
eaten by a zombie and so we justifiably ignore those other
possibilities.

The best chance that we as a civilization has of preventing a future
encounter with a zombie virus is to carefully monitor and catalog the
SNAP-contaminated planets. Although this requires the dedicated use of
the James Webb Space Telescope (JWST) to perform this task, this will
likely be insufficient to meet the challenge of monitoring all stars
with the needed signal-to-noise. Thus we strongly advocate the
construction of a fleet of no fewer than 10 JWSTs with {\em increased}
apertures (12 meters should do the trick!). These should be designed
to also operate together as a nuller interferometer so they can survey
non-transiting nearby exoplanets, which represent the main
threat. Transiting planets are generally far away, and we can all
agree that an undead neighbor is immensely more scary than a distant
zombie. Whatever the course of action, we must actively strive to
address the threat and to mitigate the risk of annihilation by an
exoplanet zombie infection.


\section*{Acknowledgements}

The authors would like to thank Vetter Brewery in Heidelberg for the
hefeweizen-fueled stagger which inspired this work. Stephen would also
like to thank Sean Raymond for not turning into a zombie and eating
his office-mate Franck without whom this work would not have been
possible.


\end{document}